\documentclass[twocolumn,amssymb, nobibnotes, showpacs, superscriptaddress, aps, prd]{revtex4}
\usepackage{graphicx,amsmath,amssymb,color}

\usepackage{soul}

\begin{document}
	
\title{Strong mesoscopic transverse growth suppression in high-speed unidirectional growth of KDP single crystals}

\author{Changfeng Fang}
\affiliation{Center for Optics Research and Engineering (CORE), Shandong University, Qingdao China 266237}
%\affiliation{State Key Laboratory of Crystal Materials, Shandong University, Jinan China 250100}
\author{Chengjie Zhu}
\affiliation{MOE Key Laboratory of Advanced Micro-Structured Materials, %School of Physics Science and Engineering, 
	Tongji University, Shanghai, China 200092}
\author{Xian Zhao}
\affiliation{Center for Optics Research and Engineering (CORE), Shandong University, Qingdao China 266237}
\affiliation{MOE Key Laboratory of Laser \& Infrared Systems, Shandong University, Jinan China 250100}
\author{L. Deng}
\affiliation{Center for Optics Research and Engineering (CORE), Shandong University, Qingdao China 266237}
\affiliation{MOE Key Laboratory of Advanced Micro-Structured Materials, %School of Physics Science and Engineering, 
	Tongji University, Shanghai, China 200092}

\date{\today}

\pacs{ky}
%\keywords{}

\begin{abstract}
	 We show a strong mesoscopic transverse growth arrest that co-exists with a high-speed longitudinal growth in single-crystalline KDP crystals with large aspect ratios. To explain this unique growth morphology, which cannot be explained by any current theories, we introduce a new set of surface concentration rate equations and demonstrate a molecular-orientation-selective surface self-shielding and channeling mechanism. We introduce a local supersaturation and calculate crystal growth driving force and growth rate, demonstrating quick arrest of both quantities as results of molecule-orientation selectivity based self-shielding and channeling effects. The growth dynamics thus derived can satisfactorily explain all experimental observations in single-crystalline KDP crystal growth reported here. 
\end{abstract}

\maketitle

\noindent {\it Introduction.} The potassium di-hydrogen phosphate ($\rm{KH_2PO_4}$ or KDP) crystal is one of the most studied optical crystals in modern optical material research \cite{booka,bookb,bookc}. Its unique electro-optic properties have made KDP crystal the primary choice for applications ranging from outdoor light shows to high-energy laser ignition for nuclear fusion \cite{app1,app2}. Because of its broad applications in fundamental science and technology extensive research has been conducted over more than a half century focusing on crystallization mechanisms and growth technologies in the pursuit of large, high quality KDP crystals. 
\vskip 5pt
\noindent  Although KDP single crystal growth processes have been studied extensively there are still remaining questions. One puzzling phenomenon is often observed crystal surface growth arrest even in aqueous solutions of favorable supersaturation. Many theories have been developed to understand key mechanisms behind this growth suppression phenomenon \cite{dz1,dz2,dz3,dz4,dz5,dz6} and it is now widely accepted that crystal growth eventually stops by one of two means: (i) depletion of growth molecules from the surrounding solution when the supersaturation reaches zero, or (ii) the presence of certain metal impurities that ``poison" the surface chemistry and impact the growth process \cite{dz6}.
\vskip 5pt
\noindent In this work, we show a strong mesoscopic KDP crystal prismatic face growth arrest that co-exists with a high-speed pyramidal face growth in a wide range of supersaturation. Currently, no theories on single-crystalline KDP growth (see discussion later) can explain this unique yet robust mesoscopic KDP crystal growth phenomenon. To explain these striking phenomena we propose a molecular-orientation-based transverse self-shielding and self-channeling mechanism. We show that this combined growth mechanisms can satisfactorily explain all experimental observations. By introducing a local supersaturation $s$, which governs the crystal growth dynamics, we show that the above described self-shielding and self-channeling mechanisms are fully compatible with the general Gibbs-Duhem thermal dynamics formalism based crystal growth driving force.    
 
\begin{figure}[htb]
	\centering
	\includegraphics[width=8 cm]{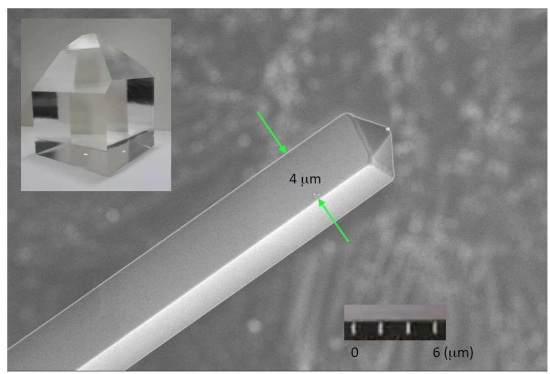}
	\caption{Morphology of a single-crystalline KDP fiber in  non-agitated growth. Growth of all prismatic faces are strong suppressed whereas the growth of the pyramidal face is abnormally fast ($> 20$ mm/day for supersaturation $\sim$ 20\%). Aspect ratio of $R>$500 can be obtained in 12 hours. Inset: Morphology of a KDP crystal grown by the conventional non-agitated process. The bulk single crystal has a low $R<$ 3 and all faces grow with a similar slow speed of a few mm/day.}
\end{figure}%fig. 0 SEM imagemolecular dynamics
\vskip 5pt
\noindent {\it Experimental results.} Figure 1 shows a typical SEM image of a high-optical quality single-crystalline KDP crystal fiber with a large aspect ratio $R$. We emphasize that the ultra-high purity aqueous solution used in our experiments is routinely used for manufacturing ultra-high grade large size KDP single crystals for high energy laser fusion applications and impurities are vigorously controlled to less than $3\times 10^{-7}$. Therefore, the impurity-based growth suppression mechanisms cannot explain experiments reported here (also see the discussion on this point later). We seed single-crystalline KDP fibers into aqueous solution and observe a strong transverse (prismatic) growth suppression in accompanying an abnormally fast longitudinal (pyramidal) growth. Typically, fibers of several cm in length can be grown in a period of 12 hours without solution agitation, which is much faster than the conventional bulk crystal growth in the same non-disturbed solution. As shown in the inset of Fig. 1, the conventional growth results in a crystal having a low aspect ratio and all surfaces, prismatic and pyramidal, grow with a slow speed of a few mm/day \cite{slow}.
\vskip 5pt
\noindent The strikingly different morphology shown Fig. 1 indicates very different growth mechanisms. One exhibits a strong transverse growth arrest whereas the other shows no growth arrest on all surfaces even though both are grown in the $\bf same$ solution. By introducing a surface molecular orientation selectivity based self-shielding mechanism we show that solute molecules with orientation favoring surface growth are effectively shielded from the surface by molecules with orientation energetically not favoring surface growth. We further show a crystal aspect ratio dependent self-channeling effect that enables a ``point effect" which strongly channels molecules from prismatic faces toward the pyramidal face of the crystal, fueling its fast growth speed and also further strengthening the self-shielding effect to prismatic faces. 
\vskip 5pt
\noindent {\it Theoretical considerations.} For simplicity of discussion, we assume that solute molecules have two orientations and we label their concentrations as $C_{\uparrow}$ and $C_{\downarrow}$, respectively. 
The growth of crystal surface requires molecules near the liquid-crystal boundary have a preferred orientation in order to achieve effective chemical bounding \cite{orref,orcal1,orcal2,orcal3} and {\it we denote this component of solute as $C_{\downarrow}$}.
\vskip 5pt
\noindent {\it I. Self-shielding effect: The role of molecular orientation.} In thermal equilibrium molecules in aqueous solution have all possible orientations. Only specifically oriented molecules have a higher probability to bond with the surface of a seed crystal due to surface chemistry and activation energy considerations. Statistically, a fraction of molecules with such a ``favorable" orientation can achieve successful anchoring/bonding on the surface, which is the first step in forming a new crystal layer. However, after quick bonding of nearby molecules with a ``favorable orientation”, a layer of molecules with ``unfavorable" orientation is left near the liquid-crystal boundary, forming an ``inert or shielding” layer. As the diffusion process continues due to the concentration gradient, molecules of both orientations reach the surface where molecules of ``favorable orientation" crystallize on the surface, but those with an ``unfavorable" orientation further accumulate in the ``shielding” layer. This process increases the thickness of the ``shielding" layer, thereby reducing the mobility and penetration probability of molecules with ``favorable" orientation, and therefore impedes the growth of the crystal surface. Consequently, crystal growth ceases. Using the terminology of atomic spin polarization, this process can be thought of as ``polarizing” material surrounding the surface, which reduces or shields the surface from further interaction. We note that this ``self-shielding" effect has similarities to ``self-poisoning" describing polymer chain crystallization processes \cite{mft,mft1}. However, the latter theory cannot explain the fast growth of pyramidal face in the presence of strong growth arrest of prismatic faces for a seed with a large $R$ as well as the slow continuous growth in all directions of a crystal of small $R$ in the {\bf same} solution. 	
\vskip 5pt
\noindent{\it II. Self-channeling effect: The role of the crystal aspect ratio.} A seed crystal with a large aspect ratio ($R>25$, defined as the length divided by the transverse dimension of the crystal fiber) has a relative flat field gradient along the long axis.  Such a surface potential can behave as an effective solute channel that creates and maintains a net movement of the solute molecules towards the pyramidal face of the seed without macroscopic solution movement. Furthermore, the low field gradient along the long axis tends to further evenly spread out “unfavorably-oriented” molecules in the ``shielding” layer along the long axis. The larger the aspect ratio, the more pronounced overall channeling effect and the more uniform distribution of the ``shielding" layer along the prismatic face \cite{width}. However, near the pyramidal face the small surface area and large curvature results in a strong local surface potential with a steep field gradient similar to the {\it point-effect}  in electrostatics, resulting in a channel-guided flow and faster growth of the pyramidal face.  This fast growth further results in the pyramidal face piercing through unevenly positioned ``shielding” fragments (due to large potential variations), exposing the pyramidal face further to the fresh environment with abundant molecules of “favorable" orientation. This self-channeling effect not only fuels high speed growth of the pyramidal face but also drains solute from the region above the ``shielding" layer near prismatic faces \cite{velocity}, further enhancing the ``shielding" effect and therefore the transverse growth arrest effect. That is, it makes the pyramidal face a strong solute collector which further depletes the supply of $C_{\downarrow}$ molecules above prismatic faces. For a seed crystal with a small $R$ ($<3$), the surface potential and field gradient are relatively similar in all directions. Therefore, no strong net directional molecule movement can be established to favorably supply molecules of "favorable" orientation to any particular surface, and all surfaces share a similar supply of solute as in the usual static bulk KDP crystal growth process, unless a macroscopic disturbance is created by an external force.   
\vskip 5pt
\noindent The combined effect of the above two mechanisms leads to a strong transverse growth arrest of prismatic faces yet a high-speed growth of the pyramidal face. In the following, we present mathematical calculations to support this molecular orientation based self-shielding and self-channeling theory. We first calculate surface adsorption energy of a molecule on the (100) plane of a KDP unit cell, revealing a strong molecular-orientation selectivity in the surface-adsorption process. We then solve rate equations for $C_{\uparrow}$ and $C_{\downarrow}$ to show the formation of a ``shielding layer". When a $C_{\uparrow}$ dependent impedance term, which characterizes the impact to the mobility of surface-heading molecules of ``favorable orientation" by the accumulating ``shielding" layer, is introduced into the rate equation of $C_{\downarrow}$, a more rapid reduction of $C_{\downarrow}$ causes surface growth to quickly cease. Correspondingly, the crystal growth driving force vanishes quickly. Finally, we solve two-component diffusion equations for a seed fiber with a large $R$ and demonstrate pyramid-face-guided solute flow and large concentration gradient, supporting the conclusion of aspect-ratio induced self-channeling effect. As a comparison we also show results from a seed crystal of small $R$, where lack of both self-channeling and local concentration gradient are demonstrated. 
\begin{figure}[htb]
	\centering
	\includegraphics[width=8.5 cm]{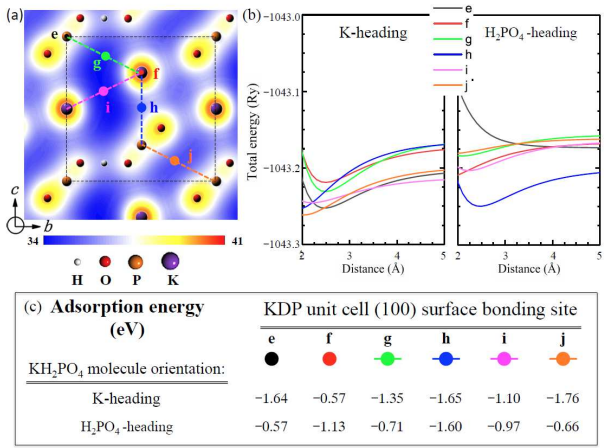}
	\caption{DFT calculations of molecule-(100)-plane interaction energy. (a) Surface energy contour and sites for surface-heading molecules. Dashed-line indicates the unit cell. (b) Total energy variations for different landing sites as a function of distance for two molecular orientations heading toward the (100) plane. (c) Sites $e$ and $j$ are most energetically favorable with strong orientation selectivity in favor of ${\rm K\!-\!heading}$.}
\end{figure}%fig.1 molecular dynamics
\vskip 5pt
\noindent {\it Numerical calculation: I. Orientation dependent surface adsorption.} The above-described molecular orientation model is based on a simplified ``two-orientation" assumption. Density functional theoretical (DFT) calculations \cite{KDP5,KDP6,KDPcal0,KDPcal1,KDPcal2,KDPcal3,KDPcal4} of realistic KDP molecule-surface interactions, which are necessarily multi-component and multi-orientation, help establish the foundation of the key concept of shielding effect by molecular orientation selectivity. Figure 2(a) shows the (100) plane surface energy of a KDP unit cell (see Supporting Information for details), which is the plane (by symmetry (010) plane as well) subject to growth arrest. Figure 2(b) shows the total energy variations for a KDP molecule of selected orientation as a function of the distance above the plane at different landing sites. It shows that ${\rm K\!-\!heading}$ to sites $e,j$ are the most energetically favorable choices to the crystal surface for effective bonding. The DFT calculated adsorption energy given in Fig. 2(c) shows strong molecular orientation selectivity by P-top and P-P-bridge sites $e$ and $j$ in favor of ${\rm K\!-\!heading}$ over ${\rm H_2PO_4\!-\!heading}$. In fact, with the exception of only K-top site $f$ all surface sites favor ${\rm K\!-\!heading}$ over ${\rm H_2PO_4\!-\!heading}$, a strong indication of overall molecular orientation selectivity. We note that DFT calculations for different ion group headings show a similar orientation preference and these results strongly support the molecular orientation model. 

\begin{figure}[htb]
	\centering
	\includegraphics[width=9 cm]{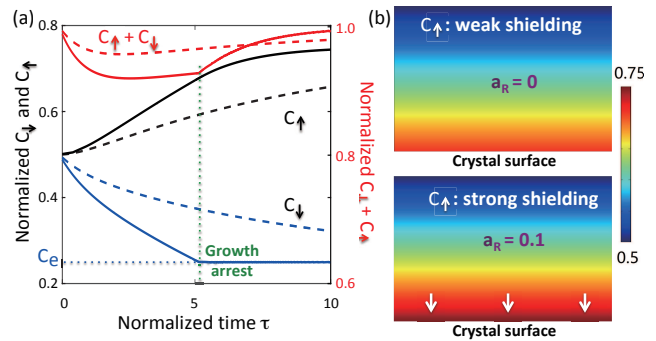}
	\caption{(a): $C_{\uparrow}$, $C_{\downarrow}$, and $C_{\uparrow}+C_{\downarrow}$ as a function of $\tau$. Solid lines: $a_R=0.1$. Dashed lines: $a_R=0$. Growth arrest sets in much earlier for $a_R=0.1$. (b): $C_{\uparrow}$ distribution contour above the crystallization plane. White arrows indicating accumulation of a ``shielding" layer above the crystal surface for $a_R=0.1$.}
\end{figure}
\vskip 5pt
\noindent{\it Numerical calculation: II. Rate equations for a two-component system.} While material diffusion does not distinguish molecular orientation surface crystallization is strongly adsorbent-orientation specific because of the requirement of energetic-selective chemical bounding. In our ``two-orientation" model, only $C_{\downarrow}$ molecules are allowed for surface bounding and therefore, crystallization. The presence of the ``shielding layer" containing $C_{\uparrow}$ necessarily impacts the mobility and impedes the penetration probability of later arriving surface-heading molecules $C_{\downarrow}$. We write dimensionless ``orientation-coupled" rate equations describing detailed surface concentration balance near the liquid-surface as \cite{bookd,booke} (for simplicity of derivation we normalize relevant quantities over a characteristic diffusion time $\tau_0$, and $\tau=t/\tau_0$), 

\begin{subequations}
\begin{eqnarray}
\frac{dC_{\uparrow}}{d\tau}&\!=\!&{\cal K}\left[C_0\!-\!\left(C_{\uparrow}\!+\!C_{\downarrow}\right)\right],\\
\frac{dC_{\downarrow}}{d\tau}&\!=\!& {\cal K}\left[C_0\!-\!\left(C_{\uparrow}\!+\!C_{\downarrow}\right)\right]\!-\!k\left(C_{\downarrow}\!-\!C_{e}\right)\!-\!a_RC_{\uparrow},\quad\quad%\\
\end{eqnarray}
\end{subequations}
where $2{\cal K}$ is the dimensionless total diffusion time constant for the solute in a diffusion-limited process. $C_0$ and $C_{e}$ are initial total concentration of solute in the aqueous solution and the equilibrium concentration of the $C_{\downarrow}$ component on the liquid-crystal surface, respectively. When $C_{\downarrow}=C_{e}$ the crystallization process ceases. The dimensionless time constant $k$ is related to the surface growth coefficient \cite{Peter} and the ``orientation-selective" surface bounding is indicated by the use of $C_{\downarrow}$ rather than $C_{\uparrow}+C_{\downarrow}$. Finally, $a_RC_{\uparrow}$ describes the impact to $C_{\downarrow}$ by the ``shielding layer" formed by $C_{\uparrow}$ near the surface. 
\vskip 5pt
\noindent Figures 3(a) and 3(b) show the impact of a non-vanishing $a_R$ to the  $C_{\downarrow}$ near the crystal surface, demonstrating a much quicker onset of the growth arrest effect. As expected, when a non-vanishing $a_R$ is introduced the molecules of ``unfavorable" orientation accumulate on the crystal surface [Fig. 3(c)], resulting in a ``shielding" layer that impedes the transport of molecules of ``favorable" orientation from the region above the surface.

\begin{figure}[htb]
	\centering
	\includegraphics[width=9 cm]{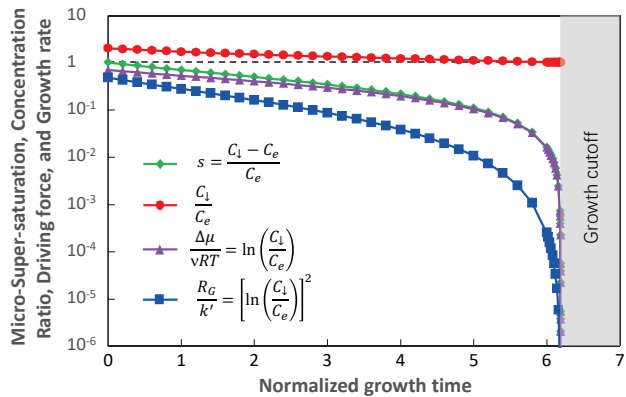}
	\caption{Plot of $s=(C_{\downarrow}-C_e)/C_e$, $C_{\downarrow}/C_e$, dimensionless crystal growth driving force $\Delta\mu/k_BT$ and growth rate $R_G/k'$ as functions of normalized diffusion time $\tau$ for $a_R=0.05$. Growth cut-off by self-shielding effect is clearly seen.}
\end{figure}
\vskip 5pt
\noindent{\it Numerical calculation: III. Crystal growth driving force.} Crystal growth is directly related to the local solute concentration. With the surface growth dynamics described by Eqs. (1a) and (1b), we can introduce a local supersaturation $s=(C_{\downarrow}-C_e)/C_e$ and in principal it is the chemical potential difference and the local supersaturation $s$, rather than the macroscopic solution supersaturation, that govern the crystal growth. Using Eqs. (1a) and (1b) the crystal growth driving force for the two-component model can be expressed as \cite{drivingforce},

\begin{equation}
\frac{\Delta\mu}{k_BT}\!=\!{\rm ln}\left(\frac{C_{\downarrow}}{C_e}\right)\!=\!{\rm ln}\left(1+s\right),
\end{equation}
where $\Delta\mu$ is the change of chemical potential, $k_B$ is the Boltzmann constant, $T$ is the absolute temperature, and $s$ is the local supersaturation near the surface.  
In Fig. 4 we plot crystal growth driving force and growth rate as functions of the normalized diffusion time $\tau$ for $a_R=0.05$ using Eqs (1a), (1b) and (2). The crystal growth driving force and growth rate quickly approach zero due to the $C_{\downarrow}$ transport impedance term.
\vskip 5pt
\noindent{\it Numerical calculation: IV. Transport of solute by the ``point-effect".} A seed crystal with a large $R$ has a pronounced ``point-effect" as in electrostatics. This creates a directional, guided flow of solute toward the sharp pyramidal face of the crystal.  This concentration flow fuels rapid growth of the pyramidal surface, resulting in it piercing through the fragmented ``shielding layer" and further exposing it to fresh materials. By drawing solute away from regions near prismatic surfaces the guided flow further enhances the transverse shielding effect described above. The diffusion dynamics is given by 
%
%\begin{subequations}
	\begin{eqnarray}
	\frac{\partial C_j}{\partial \tau}\!+\!\nabla\cdot\left(-D\nabla C_{j}\right)={\cal F}_j, \quad\quad (j=\uparrow,\downarrow), 
	\end{eqnarray}
%\end{subequations}
%  
where $D$ is the $\tau_0-$normalized diffusion constant. ${\cal F}_{\uparrow}$ (${\cal F}_{\downarrow}$) is an external force that can significantly change solute kinetics and hydrodynamics of the aqueous solution \cite{rapid1,rapid2,rapid3,rapid4}. More importantly, it can cause ``molecular-reorientation", in analogous to the optical Raman field induced atomic ``spin-reorientation" effect,  that breaks the ``shielding" layer and restarts prismatic face growth.   

\begin{figure}[htb]
	\centering
	\includegraphics[width=9 cm,angle=0]{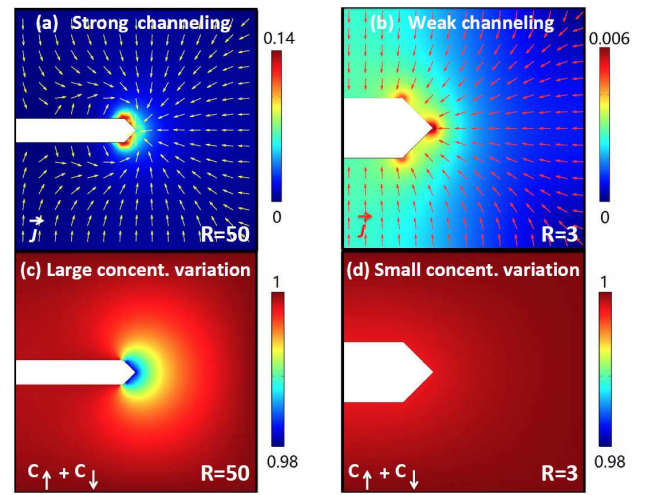}
	\caption{(a) Self-channeling by a seed with $R=50$. (b) Lack of self-channeling by a seed with $R=3$. (c) Solute is drawn strongly to the pyramidal face for $R=50$ by the ``point-effect" and a large concentration gradient. (d) Low concentration gradient near all crystal surfaces for $R=3$. }
\end{figure}
\vskip 5pt
\noindent In Fig. 5, we show solute flow (velocity) and concentration distributions of two seed crystals with different aspect ratios without ${\cal F_{{\uparrow};{\downarrow}}}$. The self-channeling effect is indicated by arrows of flow near the (100) prismatic faces toward the pyramidal face, fueling its rapid growth. The flow velocity ratio near the pyramidal face is more than 23$\times$ larger for the crystal with $R=50$ [Fig. 5(a)] over the crystal with $R=3$ [Fig. 5(b)]. The lack of solute flow on the (100) prismatic faces toward the pyramidal face in Fig. 5(b) indicates minimal solute transport on (100) faces when the aspect ratio is small. Consequently, all faces, pyramidal and prismatic, grow slowly, yielding a final crystal with a similar low aspect ratio as routinely produced in bulk crystal production. Figures 5(c) and 5(d) exhibit the solute concentration distribution near the pyramidal face for both samples. The steep concentration gradient caused by the ``point-effect" strongly drives the fast growth of the pyramidal face and at the same time further enforces the shielding effect by moving solute away from the prismatic faces.

\vskip 5pt
\noindent The strong solute flow along the prismatic surface toward the pyramidal end [Fig. 5(a)] and the substantially increased local solute gradient near the pyramidal face [Fig. 5(c)] both indicate that the driving force for the crystal growth is not the near steady-state local concentration. rather, it is the strong solute {\bf concentration gradient} that dominates the rapid growth process. Remarkably, this phenomenon resembles the very essence of stimulated light scattering by molecular normal mode vibrations \cite{Raman1,Raman2}. In analogy, we write

\begin{equation}
C_{\downarrow}(R)=C_{\downarrow}(R_p)+\sum_{m}d_{\downarrow}^{(m)}\left[\nabla^{m}C_{\downarrow}(R)\right]_{R_p}(R-R_p)^{m}.
\end{equation}
Here, $R_{p}$ is the aspect ratio measured to the pyramidal surface, and coefficient tensor $d_{\downarrow}^{(m)}$ characterizes the strength of the m-th order contribution. Equation (4) is critical for understanding highly dynamic crystalline growth processes. The crystal growth driving force for such a highly dynamic process can be expressed as \cite{Raman1,Raman2}

\begin{equation}
\frac{\Delta\mu}{k_BT}\propto {\rm ln}\left[\frac{d_{\downarrow}^{(1)}}{C_e}\nabla C_{\downarrow}\right].
\end{equation}
Indeed, the crystal growth rate measured agrees with the crystal driving force deduced from the fit of experimental data yield exactly the relation as given in Eq. (5) \cite{fit}. Physically, Eq. (5) has the similar origin as the normal mode vibration resonance enhanced polarizability. 

\vskip 5pt
\noindent It is interesting to consider the case where the pyramidal face of a KDP single crystal fiber is sealed before seeding. Under this condition the self-channeling effect is incapacitated. However, the self-shielding effect still remains and we expect to see negligible growth of all prismatic surfaces for a wide range of supersaturation. We brought a single crystal fiber ($R\sim$40) in contact with a small single crystal cube ($R\sim$1) of the similar size in the {\bf same} aqueous solution. In the next 12 hours we observed no growth of the transverse dimension of the crystal fiber and yet the small bulk crystal attached to the pyramidal surface of the crystal fiber grows steady in all directions with the usual bulk KDP crystal growth speed in the {\bf same} aqueous solution with supersaturation as high as $\sigma=25\%$. This provides the strongest evidence in support the conclusion that the transverse growth arrest is not due to any metal impurities since there is no physics reason why charged metal impurities can selectively impact the growth of the fiber section only while leave the small bulk crystal growth dynamics completely intact. 

\vskip 5pt
\noindent Finally, we stress that while the molecular-orientation-based self-shielding and self-channeling theory can satisfactorily explain all results of the present experiment it is by no means to replace many vitally important studies on crystal growth dead zone formation mechanisms. Our experiments and theoretical considerations are focused on the impact by physical environment in a mesoscopic regime that is very different from that in most molecular-level surface impurity inhibited nucleation dead zone studies. It serves as further emphasize of the complex nature of crystallization processes in different growth configurations, even for a well-studied system. 
\vskip 5pt
\noindent In conclusion, we demonstrated a strong mesoscopic transverse growth arrest that co-exists with a high-speed longitudinal growth of KDP crystals. We introduced a combined self-shielding and self-channeling mechanism which can explain all experimental observations. Our work includes three elements: (i) formation of a ``shielding layer" by molecules with ``unfavorable" orientation as the result of molecular orientation selectivity for effective surface bounding. This shielding layer impedes the mobility of molecules of ``favorable" orientation for further crystallization process. (ii) a set of surface reaction equations and a local supersaturation for crystal growth driving force. (iii) a strong self-channeled solute flow that is enabled by the ``point-effect" of crystals with large aspect ratios. This strong self-channeling effect fuels the high speed growth at the pyramidal face and also further enhances the transverse ``shielding" effect.

%\vskip 5pt
%\noindent CF and CZ contributed equally to this work. 
\vskip 5pt
%\noindent {Corresponding authors}
\noindent CF and CZ contributed equally to this work. Corresponding authors: lu.deng@email.sdu.edu.cn and zhaoxian@sdu.edu.cn. 

\vskip 5pt
\noindent {Acknowledgment:} We thank Prof. Yan Ren and Prof. Xun Sun of State Key Laboratory of Crystal Materials of Shandong University for providing Fig. 1. CF was supported by the Key Research and Development Program of the Shandong Province (No. 2018GGX102008), the China Postdoctoral Science Foundation (No.2018M632660). 
XZ was supported by the Primary Research \& Development Plan of Shandong Province (No.2017CXGC0413). CZ was supported by the National Key Basic Research Special Foundation ( No.2016YFA0302800); the Shanghai Science and Technology Committee (No.18JC1410900) and the National Nature Science Foundation ( No.11774262).

%\noindent{\bf References}

\clearpage
\centerline{\bf Supporting Information}

\subsection{Density functional computation of molecular dynamics on the (100) surface of KDP unit cell}

\noindent While the physical picture presented in this work is built upon a simplified ``two-orientation" model numerical calculations using a state-of-the-art quantum chemistry simulation package showing orientation-selective and site-selective surface bonding help support the physical reasoning of the molecular orientation hypothesis. To this end, we performed first-principle calculations using the plane-wave pseudo-potential code QUANTUM ESPRESSO \cite{DFC}. We employ Optimized Norm-Conserving Vanderbilt Pseudo-potential and PBE generalized gradient approximation functional package to treat the exchange and correlation effects. The electron wave functions are expanded in a plane-wave basis set limited by a cut-off energy of 85 Ry. Periodic boundary conditions are used on all super-cells calculations.

\vskip 5pt
\noindent Specifically, a unit cell of bulk KDP crystal is first optimized, yielding crystal lattice parameters of a = b = 7.554 $\rm{\AA}$, and c = 7.055 $\rm{\AA}$, respectively. The $x-$, $y-$, and $z-$axis in the laboratory frame correspond to the $a-$, $b-$, and $c-$crystal axis, respectively. The (100) plane is modeled by a slab of two layers in the direction normal to the plane (i.e., in the $a-$axis). A KDP molecule approaches the (100) plane along the surface norm direction, with either K-heading or ${\rm H_2PO_4}$-heading orientation. The KDP molecule is then fully relaxed on the surface to optimize the molecule-surface interaction. Multiple heading sites on the (100) plane are calculated assuming an initial 2$\rm\AA$ distance above the (100) plane [see Fig. 1a], including selected P-top (above a P atom at unit cell corner, site $e$), K-top (site $f$ near the center of the unit cell), K-K-bridge (between adjacent K atoms, site $i$), P-K-bridges (between P and K atoms near the center, sites $g,h$), and selected P-P-bridge (site $j$). The P-top site $e$ at the corner of (100) plane and the P-P-bridge site $j$ are found to be two of the most energetically favorable for surface bonding by ${\rm K\!-\!heading}$ of a KDP molecule.  These sites exhibit strong surface-heading molecule orientation selectivity, 2.877$\times$ and 2.667$\times$, respectively, in favor of ${\rm K\!-\!heading}$ for surface adsorption, as postulated by the ``two-orientation" model. In fact, all sites except the K-top site $f$ are in favor of ${\rm K\!-\!heading}$ molecular orientation for effective surface bounding. These results strong support the molecular orientation model. We also note that DFT calculations for different ion group headings show a similar orientation preference.

\subsection{Solutions to the coupled rate Equations (1a) and (1b): Self-shielding effect}

\noindent We consider the rate change of the $C_{\downarrow}$ component near the crystallization surface where we imagine an idea infinitesimal detector is located. All concentrations are normalized with respect to the total solution concentration $C_0$ before the seed crystal is introduced into the aqueous solution. We note that diffusion processes do not distinguish between molecular ``orientations" [e.g., the first term on the right side of Eqs.(1a) and (1b)]. The molecular orientation is distinguished and selected only in the bonding and crystallization process at the surface [e.g., the second and third terms on the right side of Eq. (1b)]. Rate equations (1a) and (1b) are simply the solute balance relations for relevant molecular orientation components cross an imaginary boundary plane near the crystallization surface. For instance, the first term on the right side of Eq. (1b) in the text represents the local rate increase by the concentration gradient. This term is macroscopic in nature and it is not, as a diffusion process should be, molecular orientation specific. The second term represents the amount of $C_{\downarrow}$ taken out by the crystallization process described by the dimensionless crystallization constant $k$. Notice that in the molecular-orientation theory this term is molecular orientation specific, therefore it represents a microscopic process. Here, $k$ is the time-normalized dimensionless crystallization growth constant which is related to the surface growth coefficient \cite{Peter}. Finally, the dimensionless scattering-loss parameter $a_R$ characterizes the impedance of $C_{\uparrow}$ to $C_{\downarrow}$, which is also molecular orientation dependent. The coupled rate equations (1a) and (1b) in the text can be solved analytically near the crystal surface with initial conditions $C_{\uparrow}(0)=C_{\downarrow}(0)=C_0/2$, where $C_0$ is the solution concentration before the introduction of the seed crystal. In addition, the local supersaturation $s$ can also be obtained analytically as a function of the macroscopic supersaturation $S_0$, and therefore provides much more accurate account to the crystal growth driving force and growth rate, especially when the deviation from ${\rm ln}(1+a_s/a_s^*)$ becomes significant \cite{drivingforce1}.
\vskip 5pt
\noindent When $a_R\rightarrow 0$ Eqs. (1a) and (1b) in the text recover the no-shield solutions where concentrations of both components change slowly as in the case of a bulk crystal growth under the same static crystallization conditions, as predicted by Fick's diffusion equations \cite{Peter1}. We also note that the impedance effect of the ``shielding" layer affects molecules of both orientations, and the corresponding corrections can be added to Eqs. (1a) and (1b) in the text without any difficulty.  The resulting solution will be a bit more lengthy but the main physics remains unchanged. Therefore, we did not include these corrections in the simplified rate equations (1a) and (1b).
  
\subsection{Numerical solution of the coupled diffusion equations (3): Self-channeling effect}

\noindent To understand the large aspect ratio induced self-channeling effect we solve the full coupled diffusion equation (3) given in the text numerically by taking ${\cal F}_{\uparrow}={\cal F}_{\downarrow}=0$) and by using the time-dependent surface boundary conditions given in Eqs. (1a) and (1b). We solve Eq. (3) using COMSOL Multi-Physics package and taking ${\cal K}=0.5$, $k=0.25$, and $a_R=0.1$, respectively. Other parameter combinations are also tested and the results are similar. Figure 3(a) is produced using the COMSOL Reaction Engineering interface to specify the reacting system in a perfectly mixed environment, i.e., no space dependency was assumed. Figure 3(b) and Figs. 5(a)-5(d) are generated using the COMSOL Space-Dependent Reaction Engineering interface. It utilizes the Transport of Diluted Species module and the Steady State Solver. The crystallization processes employ surface reactions boundary conditions.
 
\vskip 10pt
%\setcounter{thebibliography}{0}
%\makeatletter 
%\renewcommand{\thethebibliography}{S\@arabic\c@thebibliography}
%\makeatother

\end{document}